\documentclass[prd,tightenlines,showpacs,nofootinbib,
amsfonts,amssymb,amsmath,twocolumn,widetext]{revtex4}
\usepackage{graphicx}
\usepackage{amsmath}
\usepackage{amsfonts}   
\usepackage{amssymb}
\usepackage{url}
\usepackage{color}
\begin{document}
\newcommand{\blu}[1]{\textcolor{blue}{#1}}
\newcommand{\red}[1]{\textcolor{red}{#1}}
\newcommand{\average}[1]{\langle{#1}\rangle_{{\cal D}}}
\newcommand{\dd}{{\rm d}}
\newcommand{\etal}{{\it et al.}}
\title{The blackness of the cosmic microwave background spectrum as a probe of the distance-duality relation}

\author{George F. R. Ellis$^1$}
\email{gfrellis@gmail.com}
\author{Robert Poltis$^1$}
\email{rvpoltis@gmail.com}
\author{Jean-Philippe Uzan$^2$}
\email{uzan@iap.fr}
\author{Amanda Weltman$^1$}
\email{amanda.weltman@uct.ac.za}
\affiliation{$^1$ Astrophysics, Cosmology and Gravitation Centre, Department of Mathematics and Applied Mathematics, University of Cape Town, Rondebosch 7701, South Africa \\ $^2$ Institut d'Astrophysique de Paris, Universit\'e Pierre~\&~Marie Curie - Paris VI, CNRS-UMR 7095, 98 bis, Bd Arago, 75014 Paris, France.}

\begin{abstract}
A violation of the reciprocity relation, which induces a violation
of the distance duality relation, reflects itself in a change in the
normalisation of the cosmic microwave spectrum in such a way that
its spectrum is grey. We show that existing
observational constraints imply that the reciprocity relation cannot
be violated by more than $0.01\%$ between decoupling and today. We
compare this effect to other sources of violation of the distance
duality relations which induce spectral distortion of the cosmic
microwave background spectrum.
\end{abstract}

\date{4 January 2013}
\maketitle

In the standard cosmological model~\cite{pubook,gfrbook}, in which
the universe is described by a spatially homogeneous and isotropic
geometry of the Friedmann-Lema\^{\i}tre family, the luminosity
distance $D_L$ and angular diameter distance $D_A$ are
related by the distance duality relation,
\begin{equation}\label{e.ddr}
 D_L(z) = (1+z)^2 D_A(z),
\end{equation}
where $z$ is the redshift. This relation is actually far more
general~\cite{etherington,Ell07}. It can be shown (see
Ref.~\cite{gfrvarenna} for a demonstration) that it holds in any
spacetime in which (1) the reciprocity relation holds and (2) the
number of photons is conserved.

The {\em reciprocity relation} connects the area distances up and
down the past lightcone and is a relation between the source angular
distance, $r_{\rm s}$, and the observer area distance, $r_{\rm o}$. The
former is defined by considering a bundle of null geodesics
diverging from the source and which subtends a solid angle
$\dd{\Omega}^2_{\rm s}$ at the source.  This bundle has a cross section $\dd^2
S_{\rm o}$ at the observer and the source angular distance is
defined by the relation
\begin{equation}\label{e.s}
 \dd^2 S_{\rm o} = r_{\rm s}^2\dd{\Omega}^2_{\rm s}.
\end{equation}
Similarly, the observer area distance is defined by considering a reciprocal null geodesic bundle converging at the observer by
\begin{equation}\label{e.o}
 \dd^2 S_{\rm s} = r_{\rm o}^2\dd{\Omega}^2_{\rm o}.
\end{equation}
As long as photons propagate along null geodesics and the geodesic deviation equation holds then these two distances are related by the reciprocity relation~\cite{gfrvarenna}
\begin{equation}\label{e.reciprocity}
r_{\rm s} =  ( 1 + z ) r_{\rm o},
\end{equation}
regardless of the metric and matter content of the spacetime. While $r_{\rm o}$ is related to the angular distance, the solid angle $\dd{\Omega}^2_{\rm s}$ cannot be measured so that $r_{\rm s}$ is not an observable quantity. If one further assumes that the number of photons is conserved, the source angular distance is related to the luminosity distance, by  $D_L = (1 + z)r_{\rm s}$, which leads to the distance duality relation~(\ref{e.ddr}), where $D_A(z) = r_{\rm o}$.\\

\emph{\textbf{Varying the distance-duality relation}}: Violations of the distance duality relation~(\ref{e.ddr}) can arise from a violation of the reciprocity relation, which can occur in the case where photons do not follow 
(unique) null geodesics (e.g. in theories involving torsion and/or
non-metricity or birefringence~\cite{birefringence}) or (2) from the
non-conservation of photons, which occur e.g. when photons are
coupled to axions~\cite{axion-old} or to gravitons in an external
magnetic field~\cite{graviton-old}, to Kaluza-Klein modes associated
with extra-dimensions \cite{axion5}, or to a chameleon
field~\cite{chameleon1,chameleon2,chameleon3}. The fact that such a
violation of the distance duality relation can account for a dimming
of the supernovae luminosity~\cite{axion,axion2}, since e.g. in the case of
photon-axion mixing the luminosity distance has to be rescaled as
$D_L/\sqrt{1-P_{\gamma-a}(r)}$ with $P_{\gamma-a}$ being the
probability for a photon to oscillate in an axion after having
propagated over a distance $r$, has motivated the design of many
tests of this relation~\cite{test1,test2,test3,holanda,test4} using
independent measurements of $D_L$ and $D_A$~\cite{test1,test3} based
on the SZ effect and X-ray measurements~\cite{test2,holanda,test4}.

These tests play an important role in understanding the physics behind the acceleration of cosmic expansion~\cite{delink}. Defining the deviation from Eq.~(\ref{e.ddr}) as
\begin{equation}\label{e.eta}
 \eta(z) = (1+z)^2 \frac{D_A(z)}{D_L(z)},
\end{equation}
so the distance-duality relation holds iff $ \eta(z) =1$, it was concluded~\cite{test2} that $\eta=0.91\pm0.04$ up to $z\simeq 0.8$. \footnote{A deviation of $\eta(z)$ away from unity in optical wavelengths was studied in \cite{Avgoustidis:2010ju,Liao:2012bg} where \cite{Avgoustidis:2010ju} put strong constraints on any deviation from cosmic transparency in optical wavelengths, with $\eta (z)=1/(1+z)^\epsilon $ and $\epsilon =-0.04^{+0.08}_{-0.07}$ at $2\sigma$.} In particular, this sets strong constraints on photon-axion oscillation models~\cite{test1,test2}. It also allows us to prove~\cite{test1} that the dimming of the supernovae did not result from absorption by a grey-dust model~\cite{riess}.

It has also been pointed out~\cite{axion-cmb,axion-cmb2,axion-cmb3} in the particular case of photon-axion mixing that the oscillation probability depends on the frequency of the photon~\cite{axion-old}, so that a spectral distortion of the cosmic microwave background spectrum was expected (see below).\\

The cosmic microwave background (CMB) radiation 
is considered to enjoy one of the most precise black body spectra ever produced, that is, it has the Planck form \begin{equation} \label{bbrspect}
I_{\rm BB}(\nu,T) = \nu^3 f(\nu/T), \,\, f(\nu/T):= \frac{2h}{c^2}\frac{1}{e^{h\nu/kT}-1}
\end{equation}
where $I_{\rm BB}(\nu,T)$ is the energy per unit time (or the power)
radiated per unit area of emitting surface in the normal direction
per unit solid angle per unit frequency by a black body at
temperature $T$ at emission ($h$ is the Planck constant, $c$ the
speed of light in a vacuum and $k$ the Boltzmann constant.) We note
that the normalisation factor $\frac{2h}{c^2}$ incorporated in the
definition of $f(\nu/T)$ is crucial to the black body nature of the
spectrum, deriving directly from quantum mechanics. 

The accuracy of the CMB black body spectrum sets~\cite{mather,fixsen,BBpsec-review} constraints on various spectral distortions parametrized as the effective chemical potential $\mu$, the Compton parameter $y$ and the free-free distortion parameter $Y_{\rm ff}$ of the order
\begin{equation}
 |\mu| < 19\times 10^{-6},\quad
 |y| < 9\times 10^{-5},\quad
 |Y_{\rm ff}| < 15\times 10^{-6}.
\label{ParameterValues}
\end{equation}
This sets stringent constraints on violations of the distance duality relation induced by the non-conservation of photons, as e.g. in the case of photon-axion mixing.\\

\emph{\textbf{Changes in the CMB Spectrum}}: Our goal is to
investigate the effect of a violation
of the distance duality related to a violation of the reciprocity
relation on the spectrum of CMB photons. We assume that Eq.~(\ref{e.reciprocity}) is modified to
\begin{equation}\label{e.rec-mod}
r_{\rm s}^2=  ( 1 + z )^2 r_{\rm o}^2\beta(z),
\end{equation}
where $\beta(0)= 1$; $\beta(z)$ is a function depending on the
particular physical mechanism
responsible for the
violation of the reciprocity relation; that relation holds precisely iff
$\beta(z) = 1$.

First, the integrated flux of radiation $F$ received from an
isotropically emitting  source of intrinsic luminosity $L$ is the
amount of radiation received by the detector per unit area per unit
time, and is given by
\begin{equation}\label{e.F}
F = \frac{L}{4\pi} \frac{1}{r_{\rm s}^2 (1+z)^2}
\end{equation}
where the factor $4\pi$ arises from the integral of $\dd\Omega_{\rm s}^2$ over
the whole sky~\cite{gfrvarenna}. The specific flux $F_\nu$ received from the source,
i.e. the flux per unit frequency range, is given by
\begin{equation}\label{e.F.dF}
F_\nu \dd\nu = \frac{L}{4\pi}\frac{{\cal I}[\nu(1+z)]\dd\nu}{r_{\rm s}^2 (1+z)}
\end{equation}
where ${\cal I}(\nu)$ is the source spectrum. Note that here $\nu$ is the frequency
measured by the observer, which corresponds to a frequency $(1+z)\nu$ at emission.

What is actually measured from an extended source by a detector is
the specific intensity $I_\nu$ in a solid angle $\dd\Omega_{\rm o}^2$
in each direction of observation,
\begin{equation}\label{int1}
  I_\nu \dd\nu \equiv \frac{F_\nu \dd\nu}{\dd\Omega_{\rm o}^2}
\end{equation}
and reduces, using Eqs.~(\ref{e.F.dF}) and~(\ref{e.s}-\ref{e.o}), to
\begin{equation}
 I_\nu \dd\nu = I_{\rm s} \left(\frac{r_{\rm o}}{r_{\rm s}}\right)^2\frac{{\cal I}[\nu(1+z)]  \dd\nu}{(1+z)}
\end{equation}
where $I_{\rm s}:= L/4\pi \dd^2 S_{\rm s}$ is the source surface
brightness in that direction. It follows from Eq.~(\ref{e.rec-mod})
that
\begin{equation}\label{int55}
  I_\nu \dd\nu = I_{\rm s} \frac{{\cal I}[\nu(1+z)]\dd\nu}{(1+z)^3\beta(z)}\,.
\end{equation}
This expression is completely general (and does not assume any
specific geometry for the spacetime) and thus holds in any  curved
or flat spacetime. In the laboratory, it simply means that the
intensity of the radiation is independent of the distance from the
source, as long as the source has no relative motion compared to the
detector ($z=0$). In cosmology, the relation depends only on
redshift and is thus achromatic (so that the spectrum is redshifted
but not distorted) and is independent of area distance.

In the case of the cosmic microwave background, the photons are coupled to the electrons and baryons by Thomson scattering up to recombination [1, 2] (see \cite{Khatri:2012tv} for a description of the physical origins of the different spectral distortions). Because the collision term entering the Boltzmann
equation has a very weak dependence on the energy of the photon, the
CMB spectrum enjoys a Planck spectrum (\ref{bbrspect}). Deviations
from the Planck spectrum induced by non-linear
dynamics~\cite{stebbins,puby} are negligible as recalled in Eq. (7); see also \cite{Chluba,Khatri:2012rt, Khatri:2012tw}.

Then, denoting the emission temperature by $T_e$, we have that
\begin{equation}\label{bbrr}
I_{\rm s}{\cal I}[\nu(1+z)] = \nu^3(1+z)^3 
f\left[\frac{\nu(1+z)}{T_e}\right].
\end{equation}
Using Eq.~(\ref{int55}) and the relation $\nu\propto(1+z)^{-1}$,
which follows from the definition of redshift (which is purely a
time dilation effect, so this relation is quite independent of area
distances), we finally get
\begin{equation}\label{int66}
 I_\nu =   \frac{\nu^3}{\beta(z)} f\left[\frac{\nu(1+z)}{T_e}\right]
\end{equation}
after simplifying by a factor $(1+z)^3$.

As long as $\beta(z)=1$, the redshift prefactor combines with the
factor $\nu^3$ in Eq.~(\ref{bbrr}) so that the initial Planck spectrum
with temperature $T_e$ remains a Planck spectrum
\begin{equation}\label{int67}
 I_\nu = \nu^3f\left[\frac{\nu}{T(z)}\right] ,
\end{equation}
 with a redshifted temperature
\begin{equation}
 T(z)=\frac{T_e}{1+z}.
\end{equation}
If $\beta(z)\not=1$, then the spectrum has the form
\begin{equation}\label{spect}
I_\nu = \beta^{-1}(z) I_{\rm BB}
\left[\nu, T(z)\right].
\end{equation}
We conclude that if the reciprocity relation is violated, the
black body spectrum is observed as a grey body spectrum. However
there is no spectral distortion, so such an achromatic effect can be
confused with calibration errors.

Note that the grey body factor does not impact Wien's displacement
law, that the wavelength $\lambda_{\rm max}$ at which the intensity
of the radiation is maximum obeys
\begin{equation}
\lambda_{\rm max} = b/T
\end{equation}
where Wien's displacement constant is $b=2.8977721\times 10^{3}~
{\rm K.m}$. This corresponds to a frequency $\nu_{\rm max} = 58.8
(T/1\,{\rm K})~{\rm GHz}$. The Stefan-Boltzmann law states that the
power emitted per unit area of the surface of a black body is
proportional to the fourth power of its absolute temperature; $j_* =
\sigma T^4$, where $j_*$ is the total power radiated per unit area
and $\sigma = 5.67 \times 10^{8} {\rm m}^{2} {\rm K}^{4}$ is the
Stefan-Boltzmann constant. For a grey body, this is modified to
\begin{equation}
  j_* = \beta^{-1}(z)\sigma T^4
\end{equation}
because all intensities get changed by this same factor. Hence, the
Wien temperature \begin{equation} T_W:= b/\lambda_{\rm
max}\end{equation} and the Stefan-Boltzmann temperature
\begin{equation}T_{SB} := \{j_*\,\beta(z)/\sigma\}^{1/4}\end{equation} are
different iff $\beta(z) \neq 1.$ The first is independent of $z$, the second is not.\\

In summary, a violation of the distance duality
relation~(\ref{e.ddr}) has an imprint on the CMB spectrum. Two
possible origins of such a violation are a violation of the
reciprocity relation, or non-conservation of the number of photons.

\begin{figure}[h!]
\centering
\includegraphics[width=\columnwidth]{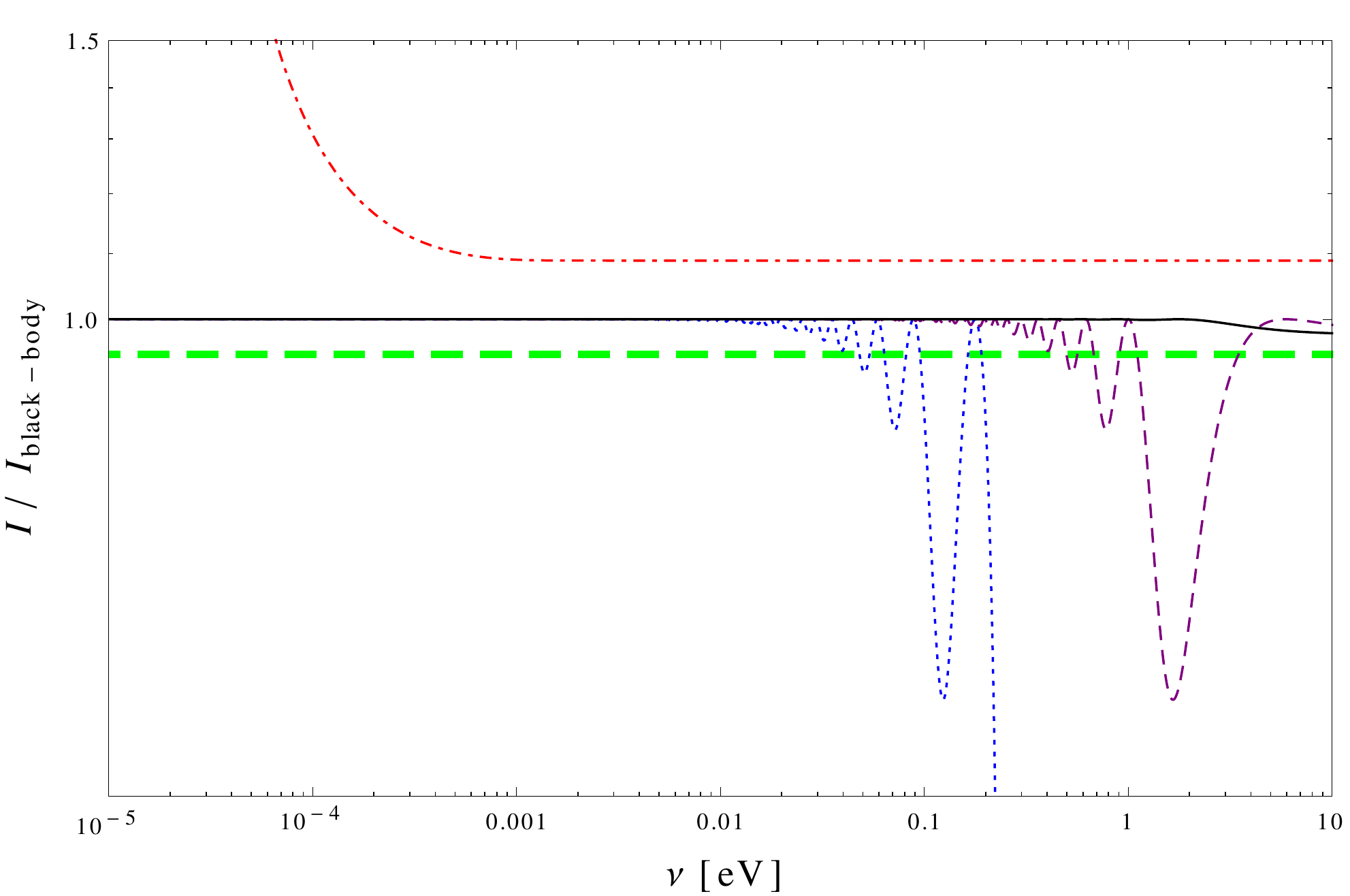}
\caption{Comparison of the function $\Phi_0(\nu,z)$ defined in Eq.~(\ref{defPhi0}) at
redshift $z=0$ for a $\mu$-type spectral distortion (red dot-dashed), a violation of the reciprocity theorem leading to a grey body (green thick dashed) and photon-axion mixing (black solid, blue dotted, and purple dashed) with different physical parameters for the distribution of the intergalactic magnetic field, calculated with the results of Ref.~\cite{axion2}. Only the solid black curve corresponds to realistic parameters.}
\label{fig1}
\end{figure}

Generically, the evolution of the distribution function $f$ of the CMB
is discussed in terms of the Boltzmann equation, ${\cal L}[f]=C[f]$.
In a Friedmann-Lema\^{\i}tre universe, the distribution function is, by symmetry,
a function of the energy $E$ and cosmic time $t$ so that the Liouville term
reduces to ${\cal L}[f]=E\partial_t f -HE^2\partial_Ef$. If the collision term
does not depend on energy, the integration of the Boltzmann equation over energy
gives the following
evolution equation for the photon number density (see e.g. Ref.~\cite{pubook,jpufluid})
\begin{equation}
\dot n+3Hn=\tilde{C}.
\end{equation}
In general, the collision term depends on energy, as for example in
the case of axion-photon mixing, so the effect of photon
non-conservation is expected to be chromatic. It follows that the
general form of the observed spectrum can be parameterized as
\begin{eqnarray}
 I(\nu,T)&=&\Phi_0(\nu,z)I_{\rm BB}\left[\nu, T(z)\right],
 \end{eqnarray}
 where
\begin{eqnarray}\label{defPhi0}
 \Phi_0(\nu,z):&=&\beta^{-1}(z) \eta(\nu,z).
\end{eqnarray}
The factor $\beta$ is related to the violation of the reciprocity
relation as shown above and the factor $\eta(\nu,z)$ to the
non-conservation of photons. In Fig~\ref{fig1} we compare the factor
$\Phi_0(\nu,z)$  for photon-axion
mixing, a violation of the reciprocity relation, and a $\mu$-type
spectral distortion. It demonstrates that each physical phenomenon
impacts a different part of the CMB spectrum.\\

Concerning the possibility of a violation of the reciprocity relation, the COBE-FIRAS experiment~\cite{mather,fixsen} showed that the CMB photons have a black body spectrum within $3.4\times10^{-8}\,{\rm ergs}.{\rm
cm}^{-2}.{\rm s}^{-1}.{\rm sr}^{-1}.{\rm cm}$ over the frequency
range from 2 to 20 cm$^{-1}$. More important concerning our work, it
showed that the deviations are less than $0.03\%$ of the peak
brightness with an rms value of $0.01\%$. This means that the
normalisation of the spectrum can be considered accurate at this
level so that it indicates a constraint of the order $ \left| \beta^{-1}(z_{\rm LSS})-1\right| < 10^{-4}$
with $z_{\rm LSS}\sim 1100$ for the redshift of the last scattering surface.\\

\begin{figure}[h!]
\centering
\includegraphics[width=\columnwidth]{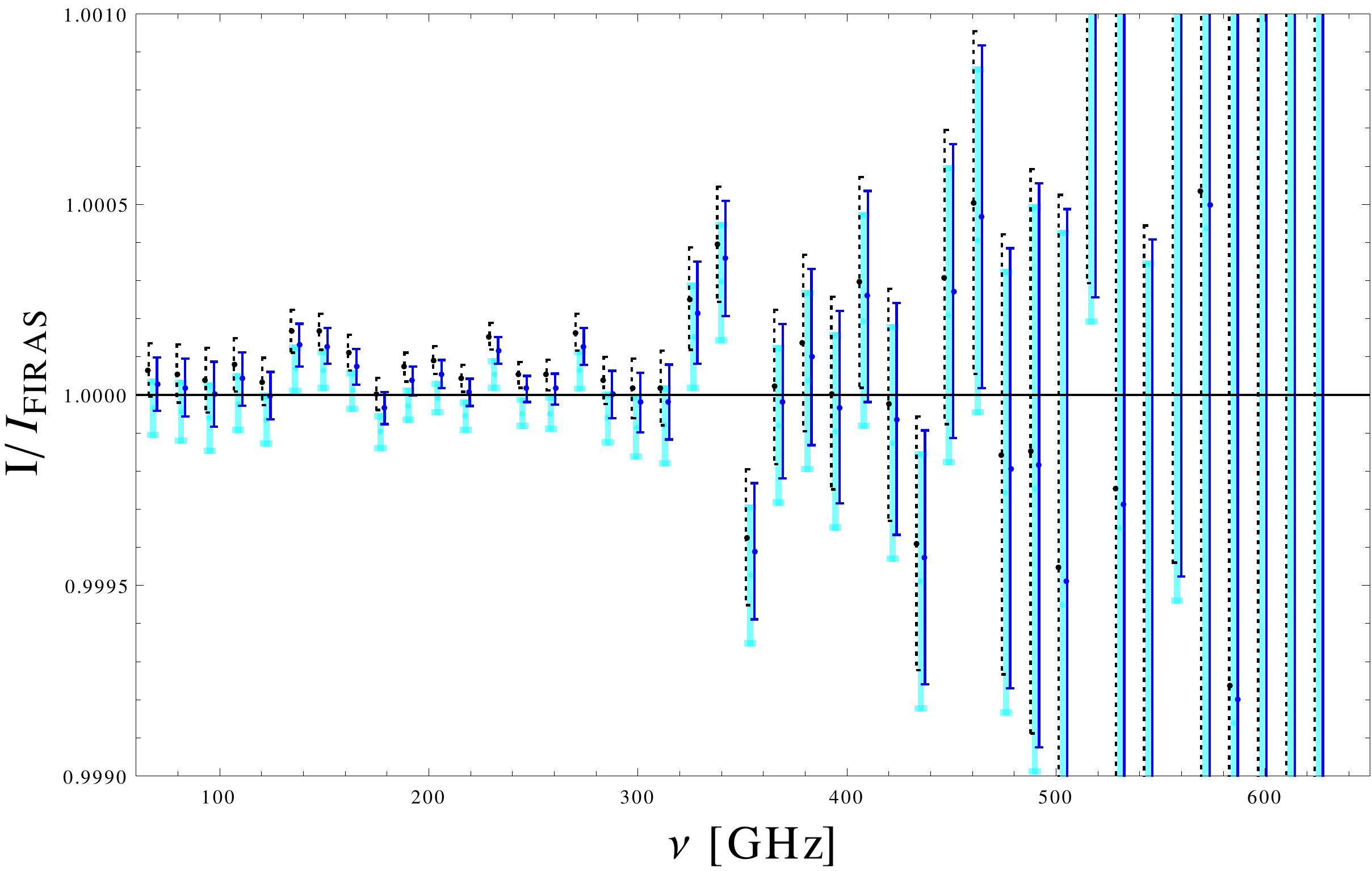}
\caption{The ratio of the spectral radiance from a $2.725$ K grey body spectrum (Eq.~(\ref{spect})) to that measured by FIRAS for $\beta^{-1}(z_{\rm LSS})-1=10^{-4}$ (black dotted), $10^{-4.2}$ (blue solid), and $0$ (cyan thick). Error bars are the $1\sigma$ uncertainty from the FIRAS data.}
\label{MonopoleComparison}
\end{figure}

To check the order of magnitude of this effect, we focus on the CMB monopole spectrum \cite{fixsen} and compare to a $2.725$K grey body spectrum for several values of $\beta^{-1}(z_{\rm LSS})$. The ratio of the expected CMB monopole signal~(\ref{spect}) to the measured CMB monopole (publicly available at \cite{FIRASsite}) is depicted in Fig.~\ref{MonopoleComparison}. Error bars designate the $1\sigma$ uncertainty of the FIRAS data. From Fig.~\ref{MonopoleComparison} we see that any violation of the reciprocity relation must be less than one part in $10^4$ at the surface of last scattering:
\begin{equation}\label{lim}
\left| \beta^{-1}(z_{\rm LSS})-1\right| < 10^{-4}.
\end{equation}
Including the spectral distortions of the CMB monopole using the constraints listed in Eq.~(\ref{ParameterValues}) (also see \cite{fixsen,Smoot:1997xt}) increases the deviation from unity of the ratio $I/I_{\rm FIRAS}$ plotted in Fig.~\ref{MonopoleComparison}, but does not significantly change any constraint on $\beta^{-1}(z_{\rm LSS})$.\\

In this article we have shown that a violation of the reciprocity
relation is associated with a deviation from blackness and that the
COBE/FIRAS data sets the constraints $\left| \beta^{-1}(z_{\rm
LSS})-1\right| < 10^{-4}$. The second factor $\eta(\nu,z_{\rm LSS})$
induces a spectral distortion and can be constrained independently;
see e.g. Refs.~\cite{axion-cmb2,axion-cmb3} for an example of the
case of photon-axion mixing. To compare to previous
constraints~\cite{test2}, we use the fact that they were based on
bolometric observations so that they concerned the parameter
$\eta(z)$ defined in Eq.~(\ref{e.eta}), that is related to
parameters introduced in this article by
\begin{equation}
 \eta(z)=\beta^{-1}(z)\frac{\int\eta(\nu,z)I_{\rm BB}\left[\nu, T(z)\right]\dd\nu}{\int I_{\rm BB}\left[\nu, T(z)\right]\dd\nu}.
\end{equation}
We recall  that $\eta=0.91\pm0.04$ up to $z\simeq 0.8$~\cite{test2}.
Note that Eq.~(\ref{lim}) gives much tighter limits over a much longer
range of redshifts.

For completeness, we shall also mention that bounds have been set on
the relation between CMB temperature and redshift, assuming a form
$T=T_0(1+z)^{1-\gamma}$. The index $\gamma$ may be constrained through measurements of the Sunyaev-Zeldovich effect for redshifts $z<1$ and fine structure excitations in quasar spectroscopy for redshifts $z>1$, specifically $\gamma=-0.004\pm0.016$ up to a redshift of $z\sim 3$. \cite{Avgoustidis:2011aa}. A non-vanishing $\gamma$ has been argued to appear in models with decaying dark energy~\cite{Lima,decayDE,Lima:1995kd}, but it has been argued to have an unphysical ansatz \cite{Khatri:2012tv}.
Such constraints on the temperature are not easily related to ours
since these analyses assume a Planck spectrum. However, when the
spectrum is no longer a Planck spectrum the different notions of
temperatures differ. We have already noted the difference between the Wien and
Stefan-Boltzmann temperatures. The latter is also related to the
bolometric temperature in the case of spectral distortions. Note
also that while the brightness and the bolometric temperatures agree
at the background and first order level, it has been
shown~\cite{pasquier} that the non-linear dynamics sources a
$y$-type spectral distortion and this would affect the brightness
and thus both the brightness and the bolometric temperatures. In
such a case~\cite{puby2} it was proposed to use occupation number
temperature defined as the temperature of a black body which would
have the same number density of photons as the actual distribution
(to be contrasted with the bolometric temperature which is the
temperature of the black body which carries the same energy density as the actual distribution).\\

\textbf{Conclusion}: We have shown that CMB spectral observations
allow one to test the distance duality relation. We have emphasized
the difference between the imprint induced by the non-conservation
of photons, usually chromatic, and a violation of the reciprocity
relation, which is achromatic. In the latter case we have shown that the CMB spectrum is a grey spectrum, with the same shape as the CMB power spectrum,
up to a normalisation factor. The FIRAS/COBE data allowed us to set
constraints of the order of $0.01\,\%$ on the relative deviation from
the reciprocity relation for the CMB.

As a final remark, we note that the limits above are for radiation coming from the surface of last scattering at $z = 1100$. However, it is likely that any effect violating the DDR relation would be cumulative, and hence proportional to distance. While it probably implies stronger constraints for closer sources, no
robust and model-independent bound can be derived. Note however that our
constraint improves those at low redshift  by at least two orders of
magnitude.
\section*{Acknowledgements}
 We thank Nabila Aghanim, Fran\c{c}ois-Xavier Desert for their insight on the CMB constraints and Cyril Pitrou for his comments.
 This material is based upon work supported financially by the National Research Foundation. Any opinion, findings and conclusions or recommendations expressed in this material are those of the authors and therefore the NRF does not accept any liability in regard thereto. G.F.R. Ellis thanks the Institut d'Astrophysique de Paris for hospitality during the early stage of this work and J.-P. Uzan thanks the Yukawa Institute for Theoretical Physics at Kyoto University, where this work was completed during the Long-term Workshop YITP-T-12-03
 on``Gravity and Cosmology 2012''.


\end{document}